\documentclass[aps,amssymb,jcp,twocolumn,showpacs]{revtex4}

\usepackage[english]{babel}
\usepackage[latin1]{inputenc}
\usepackage[dvips]{graphicx}
\usepackage{amsmath}
\usepackage{epsfig}
\usepackage{array}

\newcommand{\beq}{\begin{equation}}
\newcommand{\eeq}{\end{equation}}
\newcommand{\beqa}{\begin{eqnarray}}
\newcommand{\eeqa}{\end{eqnarray}}

\newcommand{\av}[1]{\left\langle #1\right\rangle}  %type \av{A} to make <A>

\newcommand{\I}{{\bf \mathbb I}}

\begin{document}

\title{Dynamic regimes of fluids simulated by
multiparticle-collision dynamics} 
\author{M.~Ripoll\thanks{e-mail:m.ripoll@fz-juelich.de}, 
K.~Mussawisade, R.~G.~Winkler, and G.~Gompper}
\affiliation{Institut f\"ur Festk\"orperforschung, 
Forschungszentrum J\"ulich, D-52425 J\"ulich, Germany}

\date{\today}

\begin{abstract}
We investigate the hydrodynamic properties of a fluid simulated with a
mesoscopic solvent model.  Two distinct regimes are identified, the
``particle regime'' in which the dynamics is gas-like, and the
``collective regime'' where the dynamics is fluid-like. This behavior
can be characterized by the Schmidt number, which measures the ratio
between viscous and diffusive transport.  Analytical expressions for
the tracer diffusion coefficient, which have been derived on the basis
of a molecular-chaos assumption, are found to describe the simulation
data very well in the particle regime, but important deviations are
found in the collective regime.  These deviations are due to
hydrodynamic correlations.  The model is then extended in order to
investigate self-diffusion in colloidal dispersions. 
We study first the transport
properties of heavy point-like particles in the mesoscopic solvent, as
a function of their mass and number density.  Second, we introduce
excluded-volume interactions among the colloidal particles and
determine the dependence of the diffusion coefficient on the colloidal
volume fraction for different solvent mean-free paths.
In the collective regime, the results are found to be in good
agreement with previous theoretical predictions based on Stokes
hydrodynamics and the Smoluchowski equation.

\end{abstract}

\maketitle

\section{Introduction}
\label{sec:int}

The dynamics of complex fluids such as colloidal suspensions, dilute
or semi-dilute polymer solutions, biological macromolecules,
membranes, and aqueous surfactant solutions, is often governed by the
hydrodynamic behavior of the solvent.  Due to a large separation
of length and time scales between the atomic scale of the solvent
molecules and the mesoscopic scale of the solute, direct simulation
approaches with explicit atomistic solvent are prohibitively costly 
in computer time. Therefore, several mesoscale simulation techniques 
have been developed in recent years in order to bridge the length- and
time-scale gap. In particular, lattice-gas automata (LGA)
\cite{fri86,koe90}, lattice Boltzmann methods (LBM)
\cite{mcn88,luo00,succi}, smoothed-particle hydrodynamics (SPH)
\cite{gin77,pos95b}, dissipative particle dynamics (DPD)
\cite{hoo92,esp95warren,groot97}, direct simulation Monte Carlo (DSMC)
\cite{met53,bird}, fluid particle dynamics \cite{tana00}, and others,
have been investigated. The basic idea of all these approaches is very
similar: To obtain hydrodynamic behavior on length scales much larger
than the atomic scale, the detailed interactions and dynamics of the
solvent molecules are not important; instead mass and momentum 
conservation are the essential ingredients to obtain the correct
hydrodynamic behavior. Therefore, the dynamics on the microscopic
scale can be strongly simplified, as long as the conservation laws are
strictly satisfied. The different methods listed above differ in the
way the solvent dynamics is implemented.

Two main classes of mesoscopic simulation techniques can be distinguished,
which are lattice and off-lattice methods. Lattice gas and lattice
Boltzmann methods fall into the first class, while direct simulation 
Monte Carlo, dissipative particle dynamics, and fluid particle
dynamics fall into the second class. Off-lattice approaches have the 
advantage that Galilean invariance is typically satisfied. Moreover, 
the interaction of the off-lattice solvent with 
solutes such as colloids, polymers and membranes can 
be taken into account more naturally. 

The mesoscale simulation technique, which we are investigating in
this paper, was introduced by Malevanets and Kapral \cite{kap99}
a few years ago. It is a variant of the DSMC method, in which binary
collisions are replaced by multi-particle collisions in a prescribed 
collision
volume. This method has been called multi-particle-collision dynamics 
(MPCD) or stochastic rotation dynamics (SRD). 
It employs a discrete-time dynamics with continuous
velocities and local multi-particle collisions. Mass and momentum 
are conserved quantities and it has been demonstrated that the
hydrodynamic equations are satisfied \cite{kap99,kap00}.

Certain transport coefficients, in particular the viscosity, of this 
solvent model have been studied intensively. Analytical expressions 
have been derived from kinetic theory by generalizing point-like 
collisions to finite collision volumes \cite{ihl03a,ihl03b,ihl03c,yeo03}. 
The theoretical expressions describe numerical results very well.

In this article, we study the transport coefficients as a function of
the parameters of the MPCD fluid, in particular the mean free path in
units of the size of the collision volume.  We find two distinct
regimes, in which the dynamics is either gas-like or fluid-like. This
behavior can be characterized by the Schmidt number, which measures
the ratio between viscous and diffusive transport. We find that MPCD
allows us to tune the fluid behavior such that large Schmidt numbers
are obtained and momentum transport dominates over mass
transport. Analytical expressions \cite{ihl03b,ihl03c,yeo03} for the
tracer diffusion coefficient, which have been derived on the basis of
a molecular-chaos assumption, are found to describe the simulation
data very well for large mean free paths, but fail in the fluid
regime.  The reason is a build-up of correlations among the fluid
particles by hydrodynamic interactions, which leads to enhanced
diffusion coefficients. We will show that the latter leads to
non-exponentially decaying velocity-autocorrelation functions at small
mean free paths. Independent of the mean free path, we find that the
algorithm reproduces the algebraic long-time decay typical in fluids.

In a further step, we investigate the diffusion of a heavy tracer particle 
in a MPCD solvent. It is very important to understand the contribution 
of the solvent dynamics on the solute diffusion. Two limiting 
situations are found: either Brownian or hydrodynamic behavior, 
depending on the collision time and the rotation angle. We explore 
the range of parameters where these different
dynamical behaviors appear, and show how they emerge from the mesoscopic
dynamics. 

Finally, we study self-diffusion in colloidal dispersions with
excluded-volume interactions as a function of the volume fraction.  To
this end, the MPCD method is combined with molecular dynamic
simulations. We find that such a {\em hybrid model} displays the
proper dynamics for the same parameter regime where the hydrodynamic
behavior is found for the fluid.  Our results in the
collective regime are in good agreement with previous theoretical
predictions based on Stokes hydrodynamics and the Smoluchowski equation
\cite{dhont}.

\section{The Model}
\label{sec:model}

The fluid is modeled by $N$ point particles, which are determined by 
their positions ${\bf r}_i$ and velocities ${\bf v}_i$, with
$i= 1,\ldots,N$. Positions and velocities are continuous variables, 
which evolve in discrete increments of time. The mass $m$ associated 
with the particles is taken to be the same, but more
generally, different masses can be assigned.  The algorithm consists of
two steps, streaming and collision. In the streaming step the
particles move ballistically according to their velocities during a
time increment $h$, to which we will refer as {\em collision time}. 
Thereby, the evolution rule is  
\beq 
{\bf r}_i(t+h) = {\bf r}_i(t) + h {\bf v}_i(t).  
\label{streaming.step}
\eeq
In the collision step, the particles are sorted into 
collision boxes, and interact with {\em all} other particles in the same
collision box. The collision boxes are typically the unit cells of a 
$d$-dimensional cubic lattice with
lattice constant $a$, although other geometries would be
possible. The collision is then defined as a rotation of the velocities
of all particles in a box in a co-moving frame with its center of mass. 
Thus, the velocity of the $i$-th particle after the collision is
\beq 
{\bf v}_i(t+h) = {\bf v}_{cm,i}(t) + {\cal R}(\alpha)\left[{\bf v}_i(t) -
{\bf v}_{cm,i}(t) \right],
\label{collision.step}
\eeq 
where ${\cal R}(\alpha)$ is a stochastic rotation matrix, and
${\bf v}_{cm,i}(t) = \sum_j^{(i,t)} (m {\bf v}_j)/\sum_j m$ 
is the velocity of the center of mass of all particles $j$, which are 
located in the collision
box of particle $i$ at time $t$.  The conservation of local
momentum and kinetic energy is guaranteed by construction. In two
dimensions, the rotation of the relative velocity is simply given by
an angle $\pm \alpha$. Here $\alpha$ is a parameter of the model; 
the sign is chosen randomly for each cell. In three dimensions, 
various schemes for the random collisions  
are possible \cite{kap99,all02,ihl03c}. The one employed in
this paper consist in choosing a random direction in space for each
box around which the relative velocities are rotated by an angle
$\alpha$.  A detailed explanation of the implementation 
is given in Ref.~\cite{all02}.

In order to ensure Galilean invariance for the full range of 
parameters, a {\em random shift} of the collision grid
has to be performed in the execution of the collision step 
\cite{ihl01,ihl03a}.  As a consequence of such a shift, the collision
environment of each particle is independent of the average local 
velocity, and no special reference frame exists.   
Random shifts also facilitate the transfer of momentum between 
neighboring particles. 

In the simulations, $N$ particles are initially placed at random in a
cubic system of linear extension $L$.  The average
number of particles in a collision box is $\rho = N (a/L)^d$, the
scaled number density.
Starting from an arbitrary distribution of velocities, only
a few steps are required to reach the Maxwell Boltzmann velocity
distribution. The equilibrium temperature $T$ is then 
given by the average kinetic energy $m \av{{\bf v}_i^2} = 3 k_B T$, where
$k_B$ is the Boltzmann constant. In the simulations, we scale length
and time according to $\hat x = x/a$ and $\hat t = t\sqrt{k_BT/ma^2}$,
which corresponds to the choice  
$m=1$, $a=1$, and $k_B T=1$ of reference units. The scaled mean free 
path is then given by $\lambda=\hat h$.  Basic
parameters and the definitions of dimensionless quantities are collected
in Table \ref{tab:pmt.sf}.

% Table Simple Fluid with MPCD
\begin{table}[h]
\begin{tabular}{rl}
\hline
&\underline{\sc Parameters} \\
$a$ & Collision box size  \\
$m$ & Mass of the fluid particle  \\
$T$ & Temperature  \\
$h$ & Collision time  \\
$\alpha$ & Rotation angle  \\
$L$ & Linear system size  \\
$N$ & Total number of particles \\
$\varrho$ & Mass density, \hspace{0.7cm} $\varrho = Nm/L^d$ \\
$\Lambda$ & Mean free path,  \hspace{0.8cm} $\Lambda = h \sqrt{k_B T/m}$  \\
&\underline{\sc Dimensionless quantities} \\
$\gamma$ & Decorrelation factor, \hspace{0.1cm}
$\gamma = (2/3) (1-\cos\alpha)(\rho-1)/\rho$ \\
$\rho$ & Particles per cell,  \hspace{0.7cm} 
               $\rho = \varrho a^d/m = N (a/L)^d$\\
$\lambda$ & Scaled mean free path,  \hspace{0.8cm} $\lambda = \Lambda/a$ \\
\hline 
\end{tabular}
\caption{\small Summary of relevant parameters for the simple fluid with 
the MPCD model.}
\label{tab:pmt.sf}
\end{table}

\section{Dynamical Properties}
\label{sect:visc}

The kinematic viscosity $\nu =\eta / \varrho$ has been calculated
theoretically \cite{kap99,kap00,ihl01,ihl03a,ihl03b,ihl03c,yeo03,ihl04} 
by means of kinetic theory and its
validity has been checked with simulations. The total kinematic viscosity,
$\nu = \nu_{kin} + \nu_{coll}$, is the sum of two contributions, the 
kinetic viscosity $\nu_{kin}$ and the collisional viscosity $\nu_{coll}$,
which have been calculated in two and three 
dimensions. In three dimensions, the expressions \cite{ihl03c,yeo03}
\beqa 
\label{nu_kc}
\frac{\nu_{coll}}{\sqrt{k_BTa^2/m}} &=& \frac{1}{\lambda}\frac{(1- \cos\alpha)}{18} 
     \left(1- \frac{1}{\rho}\right)  \\ 
\frac{\nu_{kin}}{\sqrt{k_BTa^2/m}} &=& \lambda \left[ \frac{1}{(4-2 \cos\alpha -2 \cos 2
\alpha)} \frac{5 \rho}{\rho-1} - \frac{1}{2}\right] \nonumber
\eeqa
have been derived.

\begin{figure}[ht]
\epsfig{file=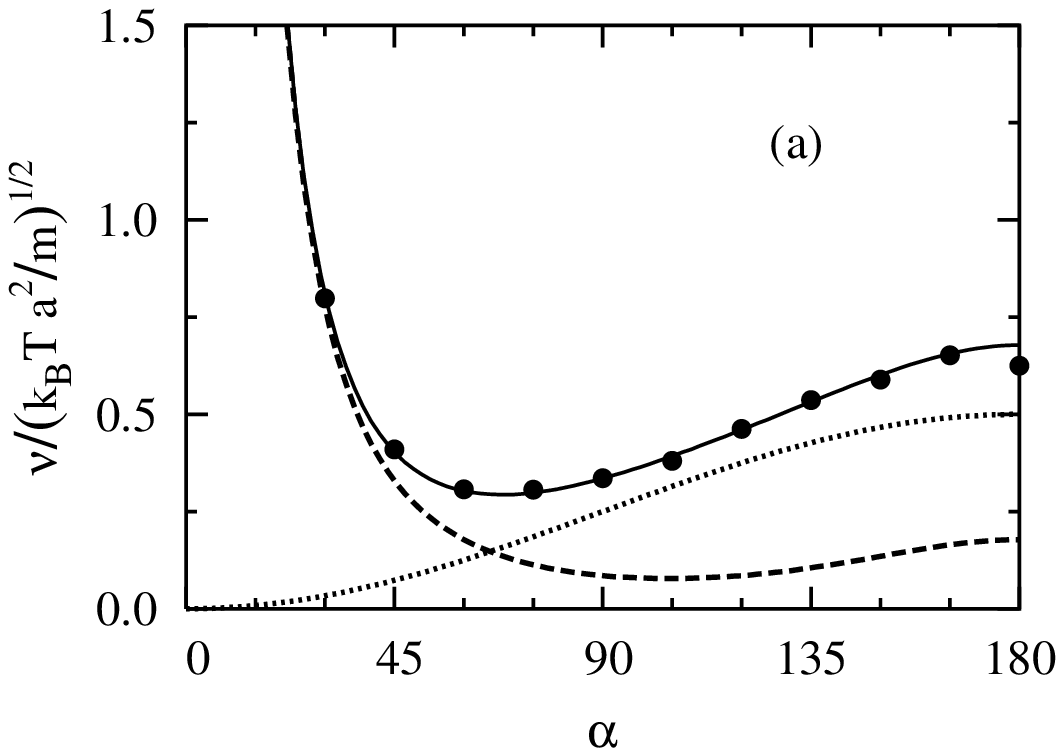,angle=-90,width=8cm}
\epsfig{file=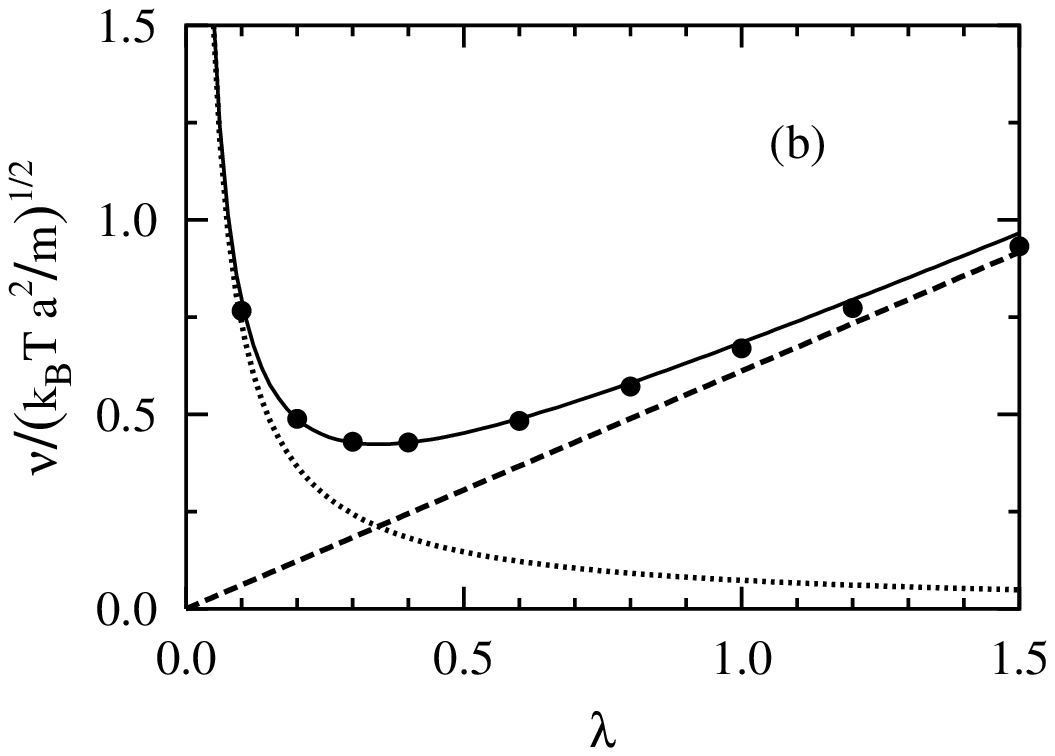,angle=-90,width=8cm}
\caption{Dimensionless kinematic viscosity for the simple fluid in
MPCD.  Symbols are the simulation results, solid line is the total
theoretical prediction, dotted line is the collisional contribution
and dashed line the kinetic contribution. In both cases the system
size is $L/a=20$. In (a) $\alpha$ dependence is displayed with
$\lambda=0.2$ and $\rho=10$. (b) shows the $\lambda$ dependence with
$\alpha=130$ and $\rho=5$.}
\label{fig:nu_kc}
\end{figure}

The total kinematic viscosity has been determined numerically by the 
procedure 
explained in Ref.~\cite{lam01}. Briefly, a three-dimensional system is 
considered with periodic boundary conditions in two dimensions and 
planar walls in the third dimension. Stick boundary conditions  
at the walls are implemented by considering
bounce-back collisions with the walls.  A
gravitational field is applied in one direction parallel to the
walls. After a relaxation time, the system reaches a stationary state 
with a parabolic velocity profile between the walls and 
in the
direction of the force. This is Poiseuille flow. It is known
\cite{tritton} that the measured maximum velocity of the parabola is
inversely proportional to the viscosity of the fluid. The viscosity
data obtained in this way are presented in Fig.~\ref{fig:nu_kc}
together with the theoretical predictions of Eq.~(\ref{nu_kc}). The
obtained agreement is quite remarkable, in contrast to the case of
other mesoscopic simulation techniques such as dissipative particle
dynamics \cite{pag98}. Density fluctuations can also be included in
the theory \cite{yeo03}, which noticeably improves the agreement with the
simulations results for small number densities; 
for $\rho=5$ and $\rho=10$, these contributions are negligible. 

\begin{figure}[ht]
\epsfig{file=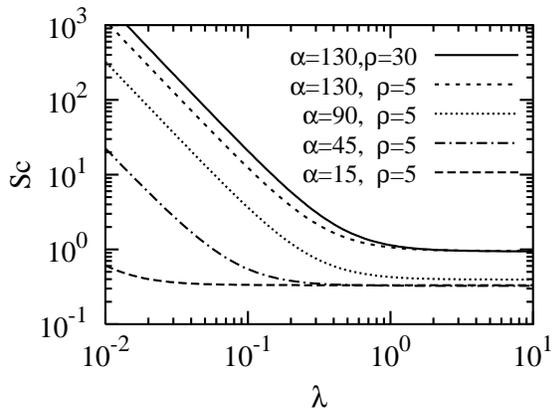,angle=-90,width=9cm}
\caption{Theoretical Schmidt number {\em versus} collision time.
The $\alpha$ and $\rho$ parameters are specified in the plot.}
\label{fig:sc}
\end{figure}

Alternative methods to determine the viscosity from simulations 
have been employed in Refs.~\cite{yeo03} and \cite{ihl03b},
where a system under shear flow and vorticity
correlations have been used, respectively.

The ratio between the kinetic and the collisional contributions to the
kinematic viscosity varies considerably with the model parameters, 
as can be seen easily from the theoretical expressions (\ref{nu_kc}).  
In Fig.~\ref{fig:nu_kc} the total kinematic viscosity and its two
contributions are plotted as a function of the rotation angle and the
collision time step. The collisional contribution is 
dominant for large collision angles and small collision times, while
the kinetic viscosity dominates in the opposite case of small 
collision angles and large collision times.

Kinetic transport is due to the movement of the
particles themselves, {\em i.e.}, when a particle moves it carries a
certain amount of the relevant quantities as momentum and energy,
while collisional transport is due to transfer of energy and momentum 
from one particle to another during collisions.  
In MPCD, kinetic transport is therefore 
dominant when the mean free path is larger than the size of the
collision box and for small values of the rotation angle. If the
rotation angle is small, there is little exchange of momentum
between particles due to collisions.  The situation where the 
kinetic transport 
dominates is characteristic for gases. 
In fluids the usual situation is the opposite, the transport 
of momentum is mainly due to collisions.

A convenient measure of the importance of hydrodynamics is the Schmidt
number $Sc=\nu/D$, where $\nu$ is the kinematic viscosity and $D$ the 
diffusion
coefficient. Thus, $Sc$ is the ratio between momentum transport and
mass transport. It is known that this number for gases is smaller than
but on the order of unity, while in fluids like water it is on the
order of $10^2$ to $10^3$. A prediction for the Schmidt number of a
MPCD fluid can be obtained from the theoretical expressions (\ref{nu_kc})
for the kinematic viscosity, and the diffusion coefficient, see
Eq.~(\ref{d.gamma}) below. In Fig.~\ref{fig:sc}, we plot the
theoretical prediction for $Sc$ as a function of the collision time
for different values of the rotation angle.  This shows that $Sc$
becomes considerably larger than unity for the same range of
parameters where the collisional viscosity is considerably larger than
the kinetic viscosity (Fig.~\ref{fig:nu_kc}). We will show that the
dynamical behavior in the two limits is fundamentally different.  We
will call the parameter region of large rotation angles and small
collision times the {\em collective regime}, and the opposite region
the {\em particle regime}. This classification has similar
consequences as the one introduced in dissipative particle dynamics
(DPD) \cite{rip01}, although we do not investigate wave-length
dependent properties here. 
%Thus, \\
%\begin{center}
%$\begin{tabular}{lcl}
%{\em Particle regime} & $\leftrightarrow$ & gas-like behavior \\
%{\em Collective regime} & $\leftrightarrow$ & fluid-like behavior 
%\end{tabular} $ 
%\end{center}

\section{Simple Fluid Correlations}

Correlations between particles are responsible for hydrodynamic 
interactions. Therefore, we are interested in characterizing the 
velocity correlations in a MPCD fluid.

\subsection{Velocity Autocorrelation Functions}
\label{sec:vacf.mc}

An analytical expression for the 
velocity autocorrelation function (VACF) has been derived in 
Refs.~\cite{ihl03b,ihl03c}. 
The collision step in Eq.~(\ref{collision.step}) can be rewritten as
\beq \label{vt.0} \begin{split}
{\bf v}_i(n h) = &{\bf v}_i((n-1)h)  + \\
&\left({\cal R}(\alpha)- \I\right) \left[{\bf v}_i((n-1)h) - {\bf v}_{cm,i}((n-1)h) \right], 
\end{split} \eeq
where $\I$ is the unit matrix and $t = n h$ is the discretized time, 
with $n$ the number of collision steps. By multiplying this expression
with the velocity at time zero and taking thermal averages, we obtain 
\beq
\label{vt.1} \begin{split} \av{{\bf v}_i(n h) {\bf v}_i(0)} =
(1-\gamma_\alpha) \av{{\bf v}_i((n-1)h){\bf v}_i(0)} \\ +
\gamma_\alpha \av{{\bf v}_{cm,i}((n-1)h){\bf v}_i(0)}.
\end{split} 
\eeq
Here, the rotational average over an arbitrary vector ${\bf A}$ 
in three dimensions is obtained from geometrical arguments to be 
\beq \label{gamma.a}
\av{\left({\cal R}(\alpha)- \I\right) {\bf A}} = - \frac{2}{3} (1 -
\cos\alpha) \av{\bf A} \equiv - \gamma_\alpha \av{\bf A}.  
\eeq 
This particular value of $\gamma_\alpha$ arises from the implementation of
the rotation chosen in this paper.  The remaining problem is 
to calculate the last term in Eq.~(\ref{vt.1}). First, we neglect
density fluctuations in the average of the center of mass velocity, 
which yields
$\av{{\bf v}_{cm,i}(nh)} \simeq \av{\sum_j^{(i,n)} {\bf v}_j} /
\rho$. Furthermore, a molecular-chaos assumption implies that 
\beq 
\av{{\bf v}_{cm,i}((n-1)h){\bf v}_i(0)} \simeq
\frac{1}{\rho}\av{{\bf v}_i((n-1)h){\bf v}_i(0)}.
\label{vcm.mc}
\eeq
This approximation means that of all the particles in the
collision box of particle $i$ after $(n-1)$ collisions, only 
particle $i$ itself makes a non-zero contribution to the
correlation function.  This is the same as assuming that none of
the other particles has any information about the state of 
particle $i$ at any time. The correlation at a certain time
step can then be expressed in terms of the previous time step as
\beq
\av{{\bf v}_i(n h) {\bf v}_i(0)} \simeq 
\left(1-\frac{\rho-1}{\rho}\gamma_\alpha \right)
\av{{\bf v}_i((n-1)h){\bf v}_i(0)}.
\label{vcm.mc1}
\eeq
This implies that in this approximation, the VACF 
shows an exponential decay,
\beq
C_v(nh) \equiv \frac{\av{{\bf v}_i(nh){\bf v}_i(0)}}
{\av{v_i^2(0)}} \simeq
(1-\gamma)^n, 
\label{cv.sf.def}
\eeq 
where the normalization factor follows from the 
equipartition theorem, $\av{{\bf v}_i^2(0)} = 3k_B T /m$. 
The decorrelation factor $\gamma$ is defined as
\beq 
\gamma = \frac{2}{3}(1-\cos\alpha) \left( 1 - \frac{1}{\rho} \right)
\equiv \gamma_\alpha \gamma_\rho. 
\label{gamma.sf}
\eeq 
>From Eq.~(\ref{cv.sf.def}), a characteristic time 
$\tau_0 = - h / \ln(1-\gamma)$ can be extracted. Up to this time, 
the VACF follows the exponential decay for every set of parameters.  
However,  the collective phenomena responsible for the
hydrodynamic behavior appear at much later times. 

\begin{figure}[ht]
\epsfig{file=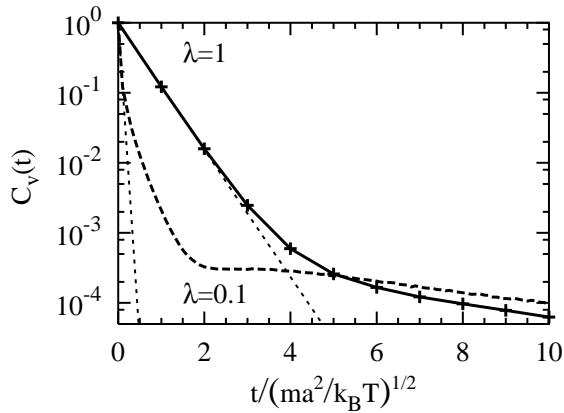,angle=-90,width=8.5cm}
\caption{Normalized velocity autocorrelation 
function as a function of the dimensionless time
for mean free paths $\lambda=1$ and  $\lambda=0.1$. 
Dashed lines correspond to the exponential decays 
in Eq.~(\ref{cv.sf.def}).  In both cases the number
density is $\rho=5$, the rotation angle $\alpha=130$, and the system
size $L/a=20$.}
\label{fig:sf_vacf}
\end{figure}

In Fig.~\ref{fig:sf_vacf}, simulation results of
the VACF are presented for two different mean free paths $\lambda$.
The theoretical prediction 
(\ref{cv.sf.def}) is also displayed for both values of
$\lambda$. 
For $\lambda=1$ the exponential decay is followed with very good
accuracy until the crossover to the long-time tail behavior occurs.
For $\lambda=0.1$ the purely exponential decay is followed only in the
first collision; for long times, a long-time-tail behavior
is observed similarly as for $\lambda=1$.  What is different in this
case is that after the first collision the system enters an
intermediate regime where the VACF decay is significantly slower than
the one described by the molecular-chaos approximation but is not yet
the algebraic tail.  Note that for the investigated rotation angle of
$\alpha=130$, the mean free path $\lambda=1$ corresponds to the
{\em particle regime}, while $\lambda=0.1$ corresponds to the 
{\em collective regime}. 

It is interesting to note that for short times, the VACF decays
monotonically only in the case that the correlation parameter $\gamma$
is smaller than unity. If $\gamma \ge 1$, Eq.~(\ref{cv.sf.def})
predicts that the VACF exhibits damped oscillations.  We have checked
that this oscillatory behavior is indeed observed in the simulations.
However, the viscosity curves show no particular features when this
happens (compare Fig.~\ref{fig:nu_kc}, where the VACF for $\rho=10$
becomes oscillatory for $\alpha\ge 132$).

\subsection{Long-Time Tails}

It is well known \cite{Alder+W,ernst71,ernst05} that the long-time behavior of 
the VACF in $d$-dimensional fluids in thermal equilibrium shows a
universal behavior. This corresponds to a power-law tail, for which
the explicit form can be calculated from a mode-coupling theory as
\cite{ernst71},
\beq \label{ltt.sf}
C_v(t) \simeq
\left(\frac{d-1}{d\rho}\right)\frac{1}{[4\pi(D+\nu)t]^{d/2}},
\eeq
where $\nu$ and $D$ are the transport coefficients of the fluid.

\begin{figure}[ht]
\epsfig{file=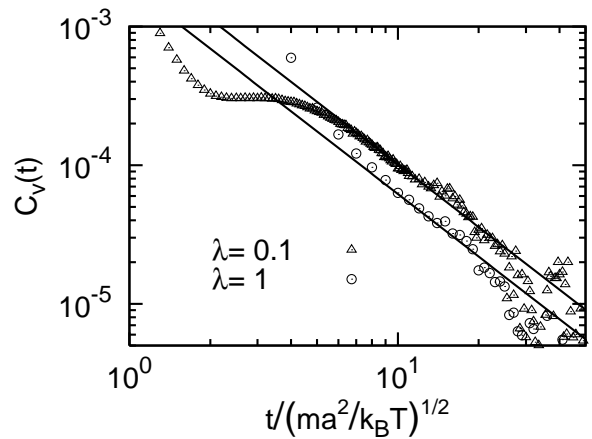,angle=-90,width=9cm}
\caption{Time dependence of the normalized velocity autocorrelation 
function.
The parameters are the same as in Fig.~\ref{fig:sf_vacf}.
The data are compared with long-time tail prediction $t^{-1/3}$. 
The amplitude predicted in Eq.~(\ref{ltt.sf}) is (within the 
statistical error) exact for $\lambda=1$ and about $10\%$  larger 
for $\lambda=0.1$.}
\label{fig:ltt.sf}
\end{figure}

The results obtained for the long-time behavior of the VACF are
consistent with the general prediction for fluids in thermal
equilibrium in Eq.~(\ref{ltt.sf}).  The algebraic power $t^{-3/2}$ is
clearly reproduced in our simulations as can be seen in
Fig.~\ref{fig:ltt.sf}.  The value of the amplitude in
Eq.~(\ref{ltt.sf}) is related to the kinematic viscosity $\nu$ and the
diffusion coefficient $D$. Since both values are known for the MPCD
fluid and discussed in this paper, quantitative comparison can also be
performed. We find that the value for $\lambda=1$ is exactly
reproduced by our simulations within the accuracy of the results,
while the amplitude obtained for $\lambda=0.1$ is about $10\%$ smaller
than the theoretic prediction. Ihle and Kroll \cite{ihl03a} obtain
good agreement in a {\em two-dimensional} MPCD fluid with the expected
$t^{-1}$ behavior over a comparable time window.

The effect of finite system size can be seen in Fig.~\ref{fig:ltt.sf}
for times $\hat t \gtrsim 20$, 
where the VACF crosses over from the algebraic to a faster, exponential
decay. This
effect is similar to that observed for the time dependence of the
temperature autocorrelation function for a random-solid 
dissipative-particle-dynamics system \cite{rip04_ltt}. There, it can 
be proved that the
correlations decay faster after a time, where hydrodynamic modes
become relevant which are truncated by the system size.

\subsection{Importance of Many-Body Correlations}
\label{sec:many_body}

In the previous section, an exponential decay of the VACF has been
theoretically predicted.
This behavior is a consequence of the approximation
in Eq.~(\ref{vcm.mc}) which neglects any correlation among the particles
in the same collision box at all times. In order to improve 
Eq.~(\ref{vcm.mc}), we have to go beyond  the molecular-chaos 
approximation. This is a formidable task. We start the procedure 
by calculating the
center-of-mass correlation average for the first collision and,
consecutively, the second and so on. 
% This is in principle possible since the velocity of a particle at a
% certain time can be expressed through Eq.~(\ref{vt.0}) in terms of the
% velocity of that particle at a previous time step and the velocity of
% the center of mass of the collision box where that corresponding
% particle was at that time.
For $n=1$ the approximation in Eq.~(\ref{vcm.mc}) is exact, $\av{{\bf
v}_{cm,i}(0){\bf v}_i(0)} = \av{{\bf v}_i^2(0)}/\rho$. This is the
reason why for the first time step, $C_v(h)$ agrees perfectly in all
simulations. For $n=2$ it reads

\begin{eqnarray}
\label{vcm.h}  
& & \av{{\bf v}_{cm,i}(h){\bf v}_i(0)} =  \nonumber \\ 
&=& \frac{1}{\rho} \sum_j ^{(i,1)} \av{\left\{{\bf
v}_j(0)+ \left({\cal R}(\alpha)- \I\right) \left[{\bf v}_j(0) - {\bf
v}_{cm,j}(0) \right] \right\}{\bf v}_i(0)} \nonumber \\ 
&=& \frac{1}{\rho}
(1-\gamma_\alpha) \av{v_i^2(0)} +\frac{\gamma_\alpha}{\rho^2}
\sum_j^{(i,1)}\sum_k ^{(j,0)} \av{{\bf v}_k(0){\bf v}_i(0)} \nonumber \\ 
&\equiv& \left( \frac{1-\gamma_\alpha}{\rho} +\frac{\gamma_\alpha}{\rho^2}
\zeta_1 \right) \av{v_i^2(0)}
\end{eqnarray}
where $\zeta_1$ denotes the number of particles that are neighbors of 
particle $i$ at both times $t=h$ and $t=0$.  
We use the term ``neighbors'' for particles within 
the {\em same} collision box.
The approximation in Eq.~(\ref{vcm.mc}) is recovered for $\zeta_1 =
1$. This is the case when only the actual particle is considered to be
in both collision boxes. As we have seen above, this is not a good
approximation in the collective regime.

We denote the average number of remaining neighbors that one particle
is revisiting after $n$ collisions as $\zeta_n$.  This number could in
principle be calculated analytically by probabilistic arguments,
but in order to get a flavor of the improvement that such
numbers produce in the theory, we determine $\zeta_n$ numerically in our 
simulations. As expected, these numbers strongly depend on the system
parameters. A detailed study has not been performed, but we have
observed that the number of remaining neighbors seems to be a universal
function of the root-mean-square displacement of the tagged particle.

\begin{figure}[ht]
\epsfig{file=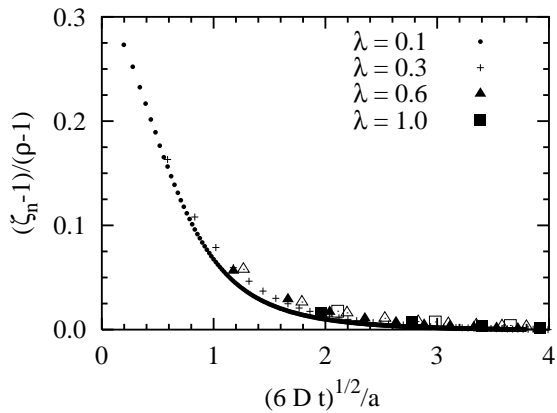,angle=-90,width=8cm}
\begin{quote}\caption{Number of remaining neighbors in a collision box 
after $n$ collisions as a function of the root-mean-square displacement. 
Solid symbols correspond to $\rho=5$, $\alpha=130$ with mean free path 
specified in the legend. Open symbols correspond to $\rho=10$, $\alpha=110$
with $\lambda = 0.6$ ({\tiny $\triangle$}) 
and $\lambda = 1$ ({\tiny $\square$}).} 
\label{fig:neigh}
\end{quote}
\end{figure}

The measured numbers $\zeta_n$ are presented in
Fig.~\ref{fig:neigh} as a function of the root-mean-square
displacement $\av{\left({\bf r}(t) - {\bf r}(0)\right)^2}^{1/2}=
\sqrt{6 D t}$, where $D$ is the diffusion coefficient and $t = n h$.
The diffusion coefficient is the one obtained from the
analytical expression which will be deduced in the next section (see
Eq.~(\ref{d.gamma})).  The data for different mean free paths seem to
fall onto a single master curve with reasonable accuracy. 
When the numerical values of $D$ (discussed in the next section) 
are used instead of the theoretical result, 
the data collapse becomes even more accurate.  
For the large mean free path $\lambda= 1$, the first collision takes
place when $\sqrt{6Dt}/a \simeq 2$, which implies that $\zeta_1 \simeq
1$ is a good approximation. Note that in the representation chosen in
Fig.~\ref{fig:neigh}, $\zeta_n \simeq 1$ corresponds to the 
abscissa.  The same displacement for a small mean
free path $\lambda=0.1$ takes place when the particle has been
involved in $80$ collisions on average. The first collision for
$\lambda=0.1$ takes place when the average displacement is much
smaller and many of the particles are still in the same
collision box, which makes $\zeta_1 \simeq 1$ a bad approximation. 
Indeed, we can infer from Fig.~\ref{fig:neigh} that 
$\zeta_1 \simeq 2.1$ for $\lambda=0.1$             
and $\rho = 5$, and $\zeta_1 \simeq 3.5$ for $\lambda=0.1$ and 
$\rho = 10$. 

Following the same procedure as employed in Eq.~(\ref{vcm.h}),
the velocity correlation function can be calculated for $n=3$,
\beq 
\label{vcm.2h}
\begin{split}
\av{{\bf v}_{cm,i}(2 h){\bf v}_i(0)} = \frac{\av{v_i^2(0)}}{\rho}
\left[ (1-\gamma)\left(1-\gamma +\frac{\gamma}{\rho}\zeta_2 \right) 
\right.\\ \left. 
+ \frac{\gamma}{\rho} (1-\gamma) \zeta_1 + \frac{\gamma^2}{\rho} 
(\zeta_2+\delta\zeta_2) \right]
\end{split}\eeq
where $\delta\zeta_2$ is determined by 
\beq
\label{eq:vacf3_delta}
\zeta_2+\delta\zeta_2 \equiv \frac{1}{\av{v_i^2(0)}}
\sum_j^{(i,2)} \sum_l^{(j,1)} \sum_k^{(l,0)} \av{{\bf v}_k(0) {\bf v}_i(0)}.
\eeq
This is the number of neighbors of particle $i$ at the two
times $t=2h$ and $t=0$ together with the neighbors of the neighbors,
or the result of ring collisions.
Let us consider two particles $i$ and $k$, which are in the same collision
box at $t=2h$ but not at $t=0$. If one, $k$, has been neighbor of a
third particle $j$ at $t=h$ and this $j$ was neighbor of $i$ at $t=0$, then
this combination also contributes to the correlation function. 
To obtain a reasonable prediction for this number is obviously not 
trivial. Furthermore, this relation
will become more interconnected and difficult to predict for further
time steps.  It can be checked that with the approximations $\zeta_n = 
1$ and $\delta \zeta_n = 0$, Eq.~(\ref{vcm.2h}) reduces to
Eq.~(\ref{vcm.mc}), and consequently the
exponential decay in Eq.~(\ref{cv.sf.def}) is recovered.

\begin{figure}[ht]
\epsfig{file=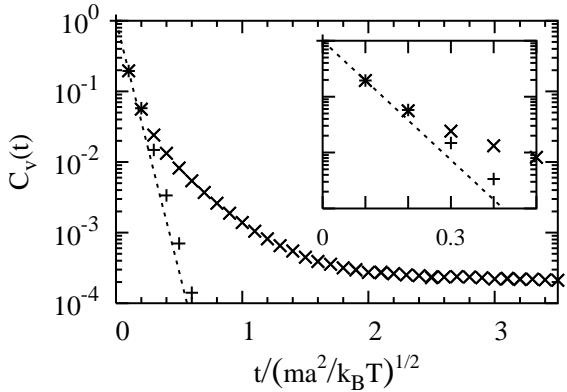,angle=-90,width=8.5cm}
\caption{Time dependence of the normalized velocity autocorrelation 
function.  The dashed line is the exponential decay in Eq.~(\ref{cv.sf.def}),
crosses ($\times$) are the simulation results and pluses (+) are the 
predicted values obtained by employing the $\zeta_n$ numbers, 
as indicated in Eq.~(\ref{eq:vacf_new}). The parameters of the 
simulation are $\rho=10$, $\alpha=110$, $\lambda=0.1$ and $L/a=20$. The inset is 
a zoom into the regime of the first few collisions.}
\label{fig:cv.neigh}
\end{figure}

Now we come back to the correlation average in Eq.~(\ref{vt.1}) which 
can be expanded with the help of Eq.~(\ref{vt.0}) --- without any 
approximation ---
\beq 
\label{eq:vacf_new}
\begin{split}
\av{{\bf v}_i(n h){\bf v}_i(0)} = \av{v_i^2(0)}(1-\gamma)^n 
\hspace{2.5 cm}\ \\ - \gamma
\sum_{k=1}^n (1-\gamma)^{n-k}\av{{\bf v}_{cm,i}((k-1)h){\bf v}_i(0)}.
\end{split}\eeq
The predictions for short times can be improved compared to
Eq.~(\ref{cv.sf.def}) by employing the results of Eqs.~(\ref{vcm.h}) and 
(\ref{vcm.2h}) on the right-hand side of Eq.~(\ref{eq:vacf_new}), but 
setting $\zeta_n \simeq 1$ and
$\delta \zeta_n \simeq 0$ for $n\ge 3$ as before.  The result is shown in
Fig.~\ref{fig:cv.neigh}. We observe that the prediction for the second
collision $C_v(2h)$ now agrees perfectly with the simulation data, 
which confirms our arguments. Nevertheless, the prediction for further
steps is still only a small improvement compared to the exponential
decay in Eq.~(\ref{cv.sf.def}). 

% The most relevant conclusion at this point is that in the collective
% regime many-body collisions are crucial for the build-up of
% correlations. This is the origin of the hydrodynamic behavior.

The most relevant conclusion at this point is that in the collective
regime the MPCD algorithm accounts for many-body collisions which are
crucial for the build-up of correlations.  This is known to be the
origin of the hydrodynamic behavior in fluids.

\section{Self-Diffusion}
\label{sect:self.dif}

We study now the consequences of the different 
behavior in the two hydrodynamic regimes, which have been introduced
in Sec.~\ref{sec:model},  on the self-diffusion coefficient. 

\subsection{Diffusion Coefficient}
\label{sec:diff.sf}

%The calculation of the mean-square displacement allows us
%to determine the value of the diffusion coefficient $D$, since for
%large times $\av{\left({\bf r}_i(t) - {\bf r}_i(0) \right)^2} \simeq 2
%d D t$.  The crossover from the ballistic to the diffusive regime
%defines a characteristic diffusion time $\tau_D \simeq 2 D m/k_B T$,
%which varies with the model parameters.  When complex fluids are
%studied, it will be important to ensure that the relevant relaxation
%times are larger than this diffusion time in order to properly take
%into account hydrodynamic effects. The diffusion time is typically
%very small (about one or two collisions) when parameter values in the
%{\em collective regime} are considered.  

In the Green-Kubo formalism, the self-diffusion coefficient is
given by $D = \frac{1}{3} \int_0^\infty dt \av{{\bf v}(t){\bf v}(0)}$. 
In the case that the time is discretized the integral has to 
be replaced by \cite{ihl01,ihl03b}
\beq
D = \frac{1}{3}  \left[ \frac{1}{2} \av{v^2(0)} h +
\sum_{n=1}^\infty \av{{\bf v}(nh){\bf v}(0)} \right] h.
\label{d.vacf}
\eeq
In order to obtain an analytical prediction for the diffusion
coefficient, an expression for $\av{{\bf v}(nh){\bf v}(0)}$ is
required. The Brownian approximation for the VACF given 
by Eq.~(\ref{cv.sf.def}) yields 
\beq
D_0 = \frac{k_B T}{m} h \left(\frac{1}{\gamma} - \frac{1}{2} \right)
\label{d.gamma}
\eeq
with $\gamma$ defined in Eq.~(\ref{gamma.sf}). This expression 
coincides with that of Ref.~\cite{ihl03b} with a different notation.
%Moreover, the same result is obtained from the 
%expression (\ref{msd.vacf}) for the mean-square displacement. 

\begin{figure}[ht]
\epsfig{file=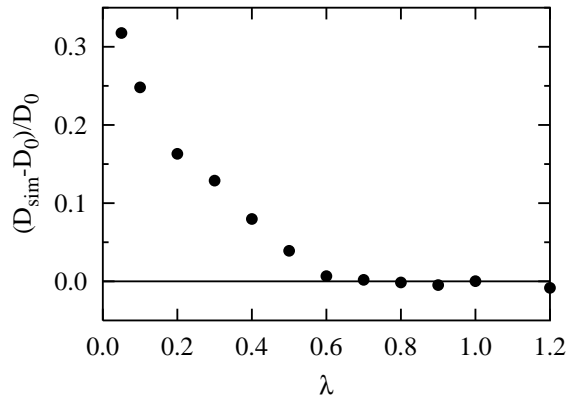,angle=-90,width=8cm}
\caption{
Relative deviation $\Delta D = (D_{sim}-D_0)/D_0$ of the simulated 
diffusion coefficient from the Brownian approximation, as a function
of the scaled mean free path $\lambda$. Full circles are simulation
results, the solid line represents the analytical 
expression of $D_0$ in Eq.~(\ref{d.gamma}).  Simulation parameters
are $\alpha=130$, $\rho=5$ and $L/a=20$.}
\label{fig:dif.coef}
\end{figure}

In the simulations, the diffusion coefficient is determined by a
linear fit of the mean-square displacement for long times. We have
checked that equivalent results for $D$ are also obtained directly
from the VACF by employing Eq.~(\ref{d.vacf}). 

Fig.~\ref{fig:dif.coef} shows the relative deviation $\Delta D =
(D_{sim}-D_0)/D_0$ of the diffusion coefficient from the expression
(\ref{d.gamma}). This expression should be a good approximation as
long as the exponential decay (\ref{cv.sf.def}) of the VACF applies.
This is indeed the case for $\lambda>0.6$, what means that the
long-time tail for these values has a negligible contribution for the
diffusion coefficient. This is reasonable since the deviation from the
exponential behavior appears when the VACF has decayed typically by
three orders of magnitude (see $\lambda = 1.0$ in
Fig.~\ref{fig:sf_vacf}).  In contrast, Fig.~\ref{fig:dif.coef} shows
that the deviation from the Brownian behavior (\ref{d.gamma}) increases
with decreasing $\lambda$ for $\lambda<0.5$.  This can be understood 
from the VACF since
for small $\lambda$ the deviation from the exponential decay
appears much earlier.  Fig.~\ref{fig:sf_vacf} shows that for $\lambda
=0.1$ the VACF has decayed only by about one order of magnitude when
the deviation starts. This translates into a noticeable increment of
the diffusion coefficient.  This difference can be understood as a
hydrodynamic enhancement of the diffusion coefficient for large values
of the Schmidt number.

%This can be understood from the VACF in Fig.\ref{fig:sf_vacf},
%since in the case of $\lambda =0.1$ the deviation
%from the exponential decay appears much earlier. This is when the VACF has

%The deviation from Brownian behavior (\ref{d.gamma}) increase with 
%decreasing $\lambda$.  This deviation is
%due to the build-up of correlations in the {\em collective regime}
%when the Schmidt number is large, as discussed for the
%mean-square displacement.

The diffusion coefficient for a simple MPCD fluid in two dimensions
has been determined by Ihle and Kroll \cite{ihl03b}.  In their
Fig.~15, results for $\lambda=0.113$ are presented as a function of
the rotation angle; deviations from the theoretical prediction are
found for large values of $\alpha$, which is in the range of
parameters which we identify as the {\em collective regime}. They
arrive at a similar conclusion that this is due to multiple
encounters among particles.  In three dimensions, some numerical
results of the diffusion coefficient have been presented in
Ref.~\cite{kap04}, and good agreement with the molecular-chaos
approximation has been found for a large range of number
densities. However, the employed parameters (which correspond to
$\lambda > 0.5$) all belong to the {\em particle regime}, where we
argue that a good agreement with the theory should be expected.

At this stage we come back to the discussion in 
Sec.~\ref{sect:visc} about the Schmidt number $Sc=\nu/D$.  The analytic
expression can be calculated from the viscosity $\nu$ in
Eq.~(\ref{nu_kc}) and the diffusion coefficient in Eq.~(\ref{d.gamma}), 
as was already pointed out in Refs. \cite{ihl03b,fal04,epl04}.
Note that $Sc$ increases rapidly for small values $\lambda \ll 1$ of the 
mean free path, where $Sc\sim h^{-2}$. This allows arbitrary
large values of the Schmidt number.  Although very small values of
the collision time significantly reduce the efficiency of the
simulations, there is a range of $\lambda$-values which are not too 
small but still display
fluid behavior corresponding to high $Sc$. On the other hand, 
the hydrodynamic enhancement of the diffusion coefficient in the
{\em collective regime} leads to values of $Sc$ which are smaller 
than predicted by the analytical approximation. By substituting the 
numerically determined diffusion coefficient, it can be checked that $Sc$ is 
indeed smaller, but still large enough to display a fluid-like behavior.

\subsection{Continuum Time Limit}

It is interesting to discuss the limit of small collision times $h \to 0$,
and small rotation angles $\alpha \to 0$.  
The leading contributions in the theoretical expressions (\ref{nu_kc}) 
of the kinetic and collisional viscosity read in this limit
\beq
\nu_{coll} \simeq \frac{m \gamma_\rho}{36 a}
\left(\frac{\alpha^2}{h}\right),
\hspace{1cm} \nu_{kin} \simeq \frac{k_B T }{a^3 \gamma_\rho} 
\left(\frac{h}{\alpha^2}\right),
\label{nu_kc0}
\eeq 
with $\gamma_\rho$ defined in Eq.~(\ref{gamma.sf}).  This result shows
that a finite viscosity is obtained in the continuum limit only if the
ratio $\alpha^2/h$ is kept constant.  The additive term due to
discrete times in Eq.~(\ref{d.gamma}) naturally vanishes in the
continuum limit, because $\gamma \sim \alpha^2$.

The expressions (\ref{nu_kc0}) for the kinetic and collisional 
contributions to the viscosity show that the collective regime,
where $\nu_{coll} \gg \nu_{kin}$, 
corresponds to $\alpha^2/h \gg 1$ in the continuum limit. In this
regime, the leading contribution to the diffusion coefficient 
(\ref{d.gamma}) is found to be
\beq
D \simeq \frac{3k_BT}{\gamma_\rho} \left(\frac{h}{\alpha^2}\right).
\label{d.gamma0}
\eeq 
The related Schmidt number
\beq
Sc = \frac{\nu}{D} \simeq \frac{1}{108}\frac{m^2 \gamma_\rho^2}{a k_B T} 
\left(\frac{\alpha^2}{h}\right)^2
\eeq
can be very large since $\alpha^2/h \gg 1$.
This shows that the model has a proper continuum limit.
However, due to the requirement of very small collision times,
this limit is not very convenient from a computational point
of view.  

It is very satisfactory to see that the Stokes-Einstein relation is 
satisfied in this case, since the diffusion 
coefficient is inversely proportional to the viscosity, 
\beq \label{stokes}
D \simeq \frac{k_B T}{ 6\pi \rho \nu_{coll} R} \ \ \ {\rm with} \ \ \  
R = \frac{2 a}{\pi \rho}
\eeq
and defines an effective particle radius inversely proportional 
to the number density. We want to emphasize, however, that the 
Stokes-Einstein relation is not only satisfied in the continuum limit,   
but always when the additive term $1/2$ in Eq.~(\ref{d.gamma}) can be 
neglected and the collisional dominates the kinetic viscosity. 
In this case, Eq.~(\ref{stokes}) is also valid.

\section{Dynamics of Embedded Particles}

After the behavior of a simple MPCD fluid has been characterized,
the next important question is
how complex fluids can be modeled. As first step, we
investigate the behavior of a single heavy point-like particle, which
could represent a solute particle or a colloidal sphere embedded in a
simple fluid. Also,  the monomers in a polymer chain can be
represented as point particles \cite{yeo00,epl04,codef04,yeo04}.  This is
a quite convenient strategy, since the solute-solvent interactions are
modeled by just including the point-like solute particles in
the collision step.  Then we study different concentrations of these
heavy particles.

\subsection{Single Heavy Tracer Particle}
\label{sec:single_heavy}

For the simulation of heavy point-like particles embedded in a 
solvent, the algorithm is the same as described for the simple fluid 
in Sec.~\ref{sec:model}. The only point where the higher mass plays a 
role is in the calculation of the velocity of the center of mass, 
where the different particle masses have to be taken 
into account via ${\bf v}_{cm,i}(t) =
\sum_j^{(i,t)} (m_j {\bf v}_j)/\sum_j m_j $.  
In thermal equilibrium, the average kinetic energy of light and heavy
particles is the same. Therefore, the average momentum of the 
heavy particle of mass $M$ is a factor $(M/m)^{1/2}$ larger than the
average momentum of a light particle. This implies that a heavy particle
has a larger contribution in the center-of-mass velocity than a 
light particle. Since the center-of-mass velocity and therefore also
the velocities of all particles after the collision step depends on
$M$ and the mass $m\rho$ of the solvent particles in a collision cell, 
the effective coupling between the solvent and the solute must depend 
in general on  the ratio $M/(m\rho)$. 

We denote the heavy particle position and velocity with capital letters
${\bf R}$ and ${\bf V}$. Of course, all types of particles are
involved in the center-of-mass calculation or other sums over 
particles.  The VACF can be calculated in the molecular-chaos 
approximation as explained in Sec.~\ref{sec:vacf.mc}, except for the 
center-of-mass correlation in 
Eq.~(\ref{vcm.mc}), which for the heavy particle yields 
\beq
\av{{\bf V}_{cm}((n-1)h){\bf V}(0)} 
\simeq \frac{M}{m\rho+M}\av{{\bf V}((n-1)h){\bf V}(0)}. 
\label{vcm.M}
\eeq
because in the collision box of the heavy particle the
total mass is $(M+m\rho)$.  The correlation at time zero depends now
on the heavy particle mass,
\beq 
\av{{\bf V}^2(0)}=3 \frac{k_B T}{M}.
\label{v0.M}
\eeq
By inserting these results in the expression equivalent to 
Eq.~(\ref{vt.1}), we obtain the molecular-chaos approximation 
for the normalized VACF of the heavy particle, 
\beq 
C_V (t)  \equiv  \frac{\av{{\bf V}(nh){\bf V}(0)}}{\av{V^2(0)}} 
\simeq (1-\gamma)^n,  
\label{vacf.M}
\eeq 
where the decorrelation factor $\gamma$ is now given by 
\beq \gamma
= \gamma_\alpha \frac{m\rho}{m\rho +M} \equiv \gamma_\alpha \gamma_1,
\label{gamma.M}
\eeq 
and $\gamma_1$ is defined for one heavy particle in the presence of
$\rho$ fluid particles, in contrast to $\gamma_\rho$ in
Eq.~(\ref{gamma.sf}), where a fluid particle is surrounded by
$(\rho-1)$ other fluid particles.

\begin{figure}[ht]
\epsfig{file=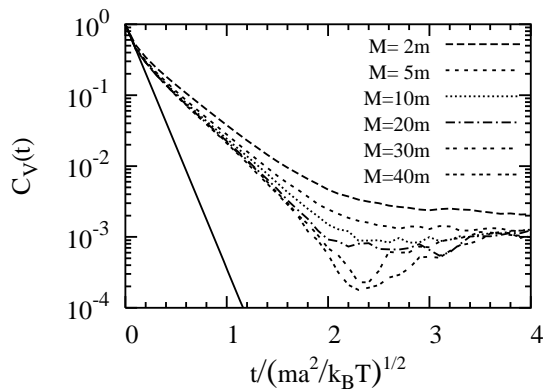,angle=-90,width=8cm}
\caption{Time dependence of the normalized velocity autocorrelation 
function for different heavy particles. Simulation parameters are
$\lambda=0.1$, $\alpha=130$, $L/a=20$ and $\rho = M/m$.  Dashed lines are
simulation results and the
solid line is the molecular-chaos approximation (\ref{vacf.M}).
}
\label{fig:vacf_M}
\end{figure}

In Fig.~\ref{fig:vacf_M} results for the normalized VACF of one heavy
particle in the collective regime are presented for different values 
of its mass.  The solvent mass density has been 
chosen to be equal to the solute mass, {\em i.e.}, $\rho = M/m$.  
In this way, $\gamma=\gamma_\alpha/2$ and the analytical expression 
(\ref{vacf.M}) is independent of the heavy particle mass.
Fig.~\ref{fig:vacf_M} shows that after the second
collision all the simulation data exhibit a non-exponential decay. 
This is not very surprising, since a similar behavior was
observed for the simple fluid in Fig.~\ref{fig:sf_vacf} for 
parameter values within the collective regime. A slightly slower decay
is displayed at lower number density $\rho$, but an asymptotic curve
is clearly approached for large values of $\rho$. The deviations
for small $\rho$ are due to the presence of density fluctuations. 

\begin{figure}[ht]
\epsfig{file=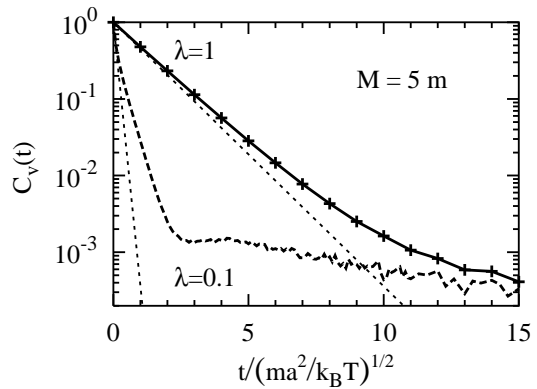,angle=-90,width= 8cm}
\caption{Time dependence of the normalized velocity autocorrelation 
function of a heavy particle of mass $M=5m$ for mean 
free path $\lambda=1$ and  $\lambda=0.1$. Dashed lines 
correspond to the exponential decays in Eq.~(\ref{vacf.M}).  
In both cases the number density is $\rho=5$ and the 
rotation angle $\alpha=130$. Compare with Fig.~\ref{fig:sf_vacf}}
\label{fig:vacf_M2}
\end{figure}

The dependence of the VACF of a single heavy tracer
particle of mass $M=5m$ on the mean free path $\lambda$ of the 
solvent is shown in Fig.~\ref{fig:vacf_M2}.
Corresponding results for the simple fluid are shown in
Fig.~\ref{fig:sf_vacf}. For $\lambda=0.1$, 
the qualitative behavior of tracer particles with $M=m$ and $M=5m$ is very
similar.
The first collision perfectly follows the molecular-chaos approximation, 
followed by a slower-than-exponential decay 
%due to the build-up of correlations 
for intermediate times and a crossover to a power-law decay 
for long times. However, note that since the exponential decay is slower 
for the heavy particle, the deviations from Brownian behavior appear 
when the VACF has decayed to approximately one third of its original 
value for the employed values of $\rho$ and $\alpha$, while for the simple 
fluid case the VACF has decayed to $6\%$ of its original value.
This implies that 
the hydrodynamic enhancement is more pronounced for particles of 
larger mass. For $\lambda = 1$, small deviations from the exponential 
decay are visible for short times; for long times, the crossover to the 
power-law behavior can be seen.

%% Mean square displacement
%% Diffusion coefficient

Analytical approximation for the diffusion coefficient 
can be calculated similar to Sec.~\ref{sec:diff.sf}. 
It reads,
\beq  
D_0 = \frac{k_B T}{M} h \left( \frac{1}{\gamma}-\frac{1}{2}\right)
\label{d_M}
\eeq 
where the decorrelation factor $\gamma$ is now given by Eq.~(\ref{gamma.M}).

\begin{figure}[ht]
\epsfig{file=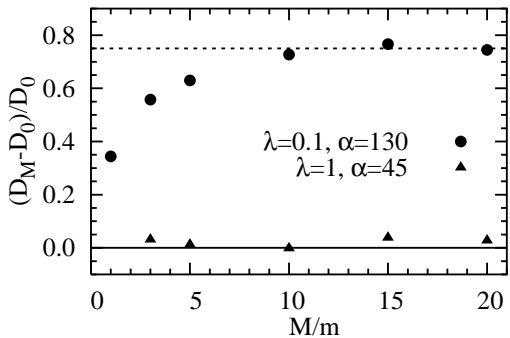,angle=-90,width= 8cm}
\caption{Relative deviation of the simulated diffusion coefficient
$D_M$ from the Brownian approximation $D_0$, in Eq.~(\ref{d_M}), 
as a function of the heavy particle mass for $\rho=5$. 
This deviations represent the hydrodynamic contribution to the 
diffusion coefficient $D_H = D_M - D_0$ in units of the Brownian 
contribution. Symbols are simulation results and the dashed 
line is a guide to the eye which represents a $75\%$ enhancement 
of the hydrodynamic term over the Brownian one.}
\label{fig:dif.M.rel}
\end{figure}

\begin{figure}[ht]
\epsfig{file=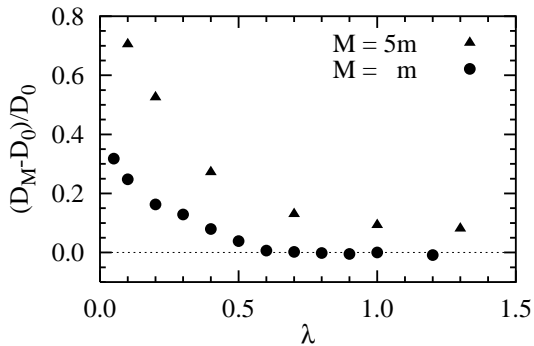,angle=-90,width= 8cm}
\caption{Hydrodynamic contribution to the diffusion coefficient in
units of the Brownian contribution as a function of the scaled mean
free path $\lambda$. Symbols are simulation measurements, and the
ordinate zero axis represents perfect agreement with the analytical
expression $D_0$. Simulation parameters are $\alpha=130$, $\rho=5$ and
$L/a=20$. Compare with Fig.~\ref{fig:dif.coef}.}
\label{fig:diff.M.h}
\end{figure}

Simulation results for the diffusion coefficient $D_M$ 
of a heavy tracer particle are plotted in Fig.~\ref{fig:dif.M.rel} 
as a function of the mass $M/m$, for fixed
solvent density $\rho=5$ and two different sets of parameters.  The
agreement of the simulations with the approximation (\ref{d_M}) is
again very good for parameter values within the particle regime,
$\lambda = 1$ and $\alpha = 45$, but not within the collective regime,
$\lambda = 0.1$ and $\alpha = 130$. This is the same behavior as observed in
the simple fluid (see Fig.~\ref{fig:dif.coef}) and indicates again the
presence of a hydrodynamic contribution to the diffusion coefficient
in the collective regime.  

In Fig.~\ref{fig:diff.M.h}, the hydrodynamic contribution to the 
diffusion coefficient (in units of the Brownian contribution) is 
plotted as a function of the scaled mean free path $\lambda$ for 
a heavy tracer particle of mass $M=5m$ and for a simple fluid 
tracer particle (compare Fig.~\ref{fig:dif.coef}). It can 
be seen that $D_H$ increases considerably for small  
$\lambda$ in both cases. This increment is significantly more pronounced 
for the heavy particle, which corresponds to the slower decay of the
VACF in  Fig.~\ref{fig:vacf_M2} for the larger mass. 
A small deviation of the VACF from the exponential decay was
observed in Fig.~\ref{fig:vacf_M2} at short times for $\lambda = 1$. 
This deviation translates into the small hydrodynamic enhancement of the 
diffusion coefficient of the heavy particle that can be seen in 
Fig.~\ref{fig:diff.M.h}, even at ``large'' mean free paths 
$\lambda\simeq 1$.

Fig.~\ref{fig:dif.M.rel} shows that for a fixed density $\rho$
in the collective regime, the hydrodynamic enhancement
increases with increasing mass of the solute particle until $M/m
\simeq 2 \rho$, and then levels off and becomes independent of the
solute mass for $M/m \gg \rho$. This is consistent with the diffusion
behavior of colloidal spheres, where the diffusion coefficient is {\em
independent} of the mass of the colloidal particles. 

Kikuchi {\em et al.} \cite{yeo03} determine numerically the friction
coefficient acting on a particle of mass $M$ and velocity ${\bf v}$ in
a MPCD solvent. Their simulation results, for a fluid of $\lambda
\simeq 0.9$, compare nicely with the analytical prediction,
independently on the mass of the particle. However, we want to point
out that this agreement is not very surprising, since their result is
obtained from the velocity
autocorrelation function after the {\em first} collision step,
where the molecular-chaos approximation is always exact (see
Sec.~\ref{sec:many_body}). 

The increase of the hydrodynamic coupling of solute and solvent with
increasing solute mass can be understood as follows.
The relative mass of the solute and solvent
particles appears in the collision step via the calculation of
the center-of-mass velocity.  If solute particles have  
the same mass as solvent particles and there is a large number of
solvent particles per cell, the solvent particles transfer a large
random momentum to the solute particle.  Simultaneously, the effect of
the solute particle momentum on the solvent is small.  For this
reason, the hydrodynamic contribution to the diffusion constant of a
particle of equal mass, shown in Fig.~\ref{fig:dif.coef}, is only of
the order of $30\%$ for the largest Schmidt number considered.  In
contrast, this hydrodynamic enhancement is $65\%$ when $M/m \simeq
\rho$ and $75\%$ when $M/m \gtrsim 2 \rho$, as can be seen in
Fig.~\ref{fig:dif.M.rel}.  A very large mass of the solute particle is
not very convenient either, because it implies a large ballistic
regime and a long diffusion time.  Therefore, we conclude that a mass
$M/m \simeq \rho$ for the solute particle is a optimal choice to
enhance the hydrodynamic coupling between solute and fluid particles.

\subsection{Finite Concentration of Heavy Point-Like Particles}
\label{sec:heavy_dispersion}

At a finite concentration of solute particles, an important question is 
to which extent solute particles build up 
hydrodynamic interactions among themselves through the fluid particles
when simulated with MPCD. We study therefore systems with different
concentrations of heavy particles for sets of parameters within the
particle and the collective regimes, respectively.
We address this question by investigating the tracer-diffusion
coefficient.

\begin{figure}[ht]
\epsfig{file=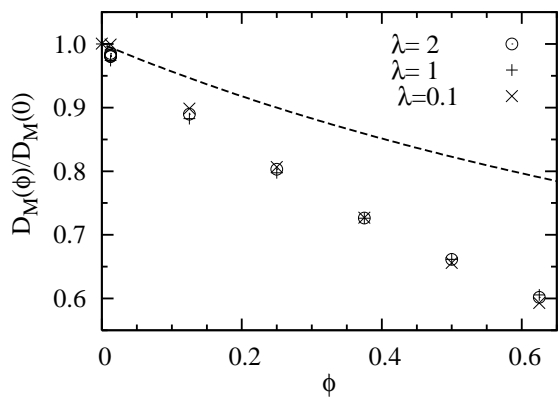,angle=-90,width= 8cm}
\caption{Diffusion coefficients for a heavy particle as a function of
the concentration $\phi = N_M (a/L)^3$, normalized with the diffusion
coefficient $D_M(0)$ of heavy particles at zero number density.  The
dashed line is the analytical approximation from Eqs.~(\ref{d_M}) and
(\ref{gam.phi}), symbols correspond to the simulation data with
$\lambda$ specified in the legend. The other parameters are 
$M/m=\rho=5$, $\alpha =130$ and $L/a=20$.}
\label{fig:d_nm}
\end{figure}

Simulations with different heavy particle concentrations are performed
by changing the total number $N_M$ of heavy particles but keeping
fixed the volume $V=L^3$ and the number $N$ of solvent particles. The
corresponding number density of heavy particles is defined as
$\phi=N_M(a/L)^3$.  In Fig.~\ref{fig:d_nm}, the diffusion coefficients
for three different values of the mean free path are displayed. 
Very surprisingly, when the data are normalized by the corresponding
diffusion coefficients in the limit of vanishing density $\phi$,
all three data sets, which are both in the particle
and the collective regime, collapse onto a single curve. We 
recall that the hydrodynamic enhancement for the diffusion coefficient
of a single heavy particle, here denoted as $D_M(0)$, is quite
different among these three values of $\lambda$ (see Fig.~\ref{fig:diff.M.h}). 
It can be inferred from the data collapse in Fig.~\ref{fig:d_nm} that 
there is no extra hydrodynamic contribution among these heavy particles,
which is consistent with the idea that there is no hydrodynamic 
screening for point particles \cite{ahl01,ladd96}. 

The dependence of the diffusion coefficient on the heavy particle number
density can be understood along the same lines as for the simple fluid
or the single heavy particle. We assume that in each collision box
there is a fixed number of fluid particles $\rho$, but that the
number of heavy particles $n$, fluctuates from one collision box to
another.  The probability $P(n)$ of a given heavy particle to be found in
a cell with a total of $n-1$ other heavy particles is given by the
Poisson distribution function, $P(n) = e^{-\phi} \phi^{n-1}/(n-1)!$.  
The corresponding decorrelation factor for a  heavy particle in a
collision box with $(n-1)$ other heavy particles and $\rho$ fluid ones is
\beq 
\gamma_n=1-M/(\rho m + n M),
\eeq 
compare the definition of $\gamma_1$ in Eq.~(\ref{gamma.M}) for a 
single heavy particle in a collision box. The diffusion coefficient 
is then given by Eq.~(\ref{d_M}), where the decorrelation factor is now
$\gamma = \gamma_\alpha \sum_{n=1}^{\infty} P(n) \gamma_n$. 
In the regime of low number density, $\phi \ll 1$, this implies
\beq
\gamma = \gamma_\alpha \left[(1-\phi)\gamma_1 
+ \phi \gamma_2   + {\cal O}\left(\phi^2\right)\right].
\eeq
In the special case of $\rho=M/m$, the sum can be evaluated analytically
and yields   
\beq \label{gam.phi}
\gamma =  \gamma_\alpha \left( 1 - \left(
e^{-\phi} + \phi -1  \right)/\phi^2\right).
\eeq 

In Fig.~\ref{fig:d_nm} the simulation data for the normalized
diffusion coefficient at different volume fractions are compared with
the theoretical prediction obtained from Eq.~(\ref{d_M}) with the
decorrelation function in Eq.~(\ref{gam.phi}). It can be seen that
this prediction overestimates the values for the diffusion coefficients.
Further studies are required to understand the origin of this deviation.

\section{Hybrid Dynamics}

In order to go one step further in the development of an efficient 
simulation technique for suspensions of colloidal particles with MPCD, 
we next investigate the
effect of excluded-volume interactions between the heavy particles.
To this end, the MPCD algorithm has to be combined with standard 
{\em molecular dynamics} (MD) for the solute particles.

\subsection{The Model}

We consider a dispersion of spherical colloidal particles in three
dimensions.  The interactions of solvent particles among themselves
and with colloids take place in the MPCD collisional step, exactly in
the same way as described for the heavy point-like particles in
Sec.~\ref{sec:single_heavy}.  However, the streaming step
(\ref{streaming.step}) is used only for the solvent particles.  The
position update of the colloidal particles is performed in several MD
steps between MPCD collisions.  In these MD steps, colloids interact
via an excluded-volume potential. We use the truncated repulsive
Lennard Jones potential \cite{slj76}
\beq \label{RLJ} 
V^{RLJ}(r) = \left\{\begin{tabular}{rcl}
$4\varepsilon\left[\left(\frac{\sigma}{r}\right)^{12}
-\left(\frac{\sigma}{r}\right)^6\right] + \varepsilon$, && $r \le
r_{min}$ \\ $0$, && $r > r_{min}$
\end{tabular} \right.
\eeq 
where $r$ is the distance between the centers of the colloidal
particles. The parameter $\sigma$ is related to the particle diameter; 
it is  
chosen to equal the collision box length, $\sigma=a$, so that there is
typically no more than one colloid particle in each collision box. The
potential strength is taken to be equal to the thermal energy,
$\varepsilon=k_B T$,  the cut-off radius is
$r_{min}=2^{1/6}\sigma$, and the mass of the particles is taken to be 
$M=5m$. The MD time steps are integrated with the velocity-Verlet 
algorithm \cite{allen} with a time step 
$\Delta t = 0.002 \sqrt{\varepsilon/ma^2}$.

In other words, we consider a system of colloidal particles
interacting through repulsive Lennard Jones potentials whose positions
and velocities evolve in  discrete time intervals $\Delta t$.
This procedure is interrupted every $h/\Delta t$ steps for the
interaction with the fluid particles. This interaction is a MPCD event
where solvent and solute particles interchange momentum.  This
implies that the solvent particles can enter the cores of the
colloidal particles, but the colloids cannot interpenetrate each
other.

The hybrid model described here is a variant of the model 
introduced previously by Malevanets and Kapral \cite{kap00,kap_book.1}. 
In their model, both the solute-solute and solute-solvent interactions
were taken into account through excluded-volume potentials
with MD, and only the solvent-solvent interactions were
mesoscopically described through MPCD. 
The advantage of the model described here comes from the fact that in
the MD steps just the solute particles are considered.  This leads
to a considerable speed up of the simulations.

\subsection{Diffusion in Colloidal Dispersions}

We measure the diffusion coefficient of the dispersion through the 
mean-square displacement of a tracer particle, as before. Simulations 
are performed for different colloidal concentrations.  The volume 
fraction of colloidal particles $\varphi$ is 
the fraction of the total volume $V$ occupied by the colloidal particles,
$\varphi= (\pi/6)\sigma_{eff}^3\rho_M$, where the effective diameter
$\sigma_{eff}$ is determined by the Barker-Henderson expression
\cite{barker_hend}
\beq
\sigma_{eff} = \int_0^{r_{min}} dr \left[1-\exp(-V^{RLJ}(r)/k_BT)\right].
\eeq
For our choice of Lennard-Jones parameters, this gives 
$\sigma_{eff} = 1.01 \sigma$.  The number density of colloidal 
particles is $\rho_M=(N_M-1)/V\simeq N_M/V$, where $N_M$is the number of 
heavy particles with excluded-volume interactions.

%%  MD simulations - Gas 

\begin{figure}[ht]
\hspace{-1cm}\vspace{-0.5cm}
\epsfig{file=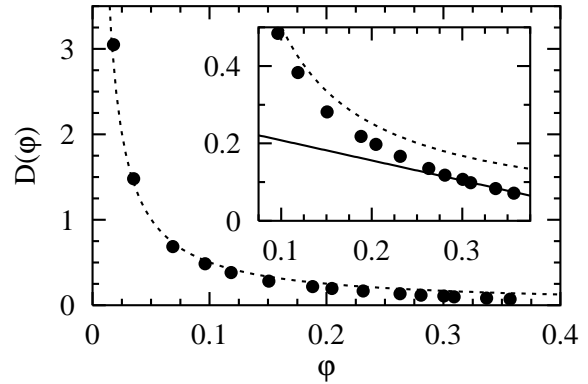,angle=-90,width=9 cm}
\caption{Diffusion coefficient for colloidal particles without
solvent as a function of the volume fraction $\varphi$. 
Symbols are simulation results, the dashed line corresponds
to the analytical prediction (\ref{diff.md}). The inset is a zoom 
over the small  values of the diffusion coefficient, and the 
solid line is a linear extrapolation for large values of $\varphi$.}
\label{fig:coll_md}
\end{figure}

For later comparison and better understanding of our hybrid
model results, we recall first the basic behavior of a system with
excluded-volume interactions only. In Fig.~\ref{fig:coll_md} we show
the results for the diffusion coefficient of a MD simulation of 
repulsive  Lennard-Jones particles. Kinetic theory  
for hard spheres predicts in the low-density limit \cite{yip}
\beq \label{diff.md}
D_{\rm{\tiny MD}}(\varphi) = 
\frac{3}{8 \sigma_{eff}^2}\sqrt{\frac{k_B T}{\pi M}} \frac{1}{\rho_M},
\eeq
This analytical prediction is depicted in Fig.~\ref{fig:coll_md} 
together with the simulation results. It can be seen that 
for small volume fraction, the  $\varphi^{-1}$ behavior
is properly reproduced, while for large volume fractions a 
linear behavior can be inferred.

%% MPCD simulations

The density dependence of the self-diffusion coefficient 
of colloidal hard spheres in a hydrodynamic bath has been
calculated in Ref.~\cite{dhont},   
\beq \label{dcoll.hyd}
D_S(\varphi) = D_S(0) \left[ 1 - 2.1 \varphi + 
{\cal O}\left(\varphi^2\right)\right].
\eeq 
The diffusion coefficient now decreases linearly with
the volume fraction, in contrast with the kinetic theory result 
(\ref{diff.md}) for a gas of hard spheres.  In the calculation of
Eq.~(\ref{dcoll.hyd}), Brownian and hydrodynamic terms have to be
considered, and it has been found that the hydrodynamic terms
almost cancel. For a colloidal dispersion in a Brownian bath 
\cite{dhont} the first-order correction in Eq.~(\ref{dcoll.hyd}) 
equals $-2.0\varphi$.  Thus, no significant differences are
expected between Brownian and hydrodynamic measurements of the
diffusion coefficient. 
 
\begin{figure}[ht]
\epsfig{file=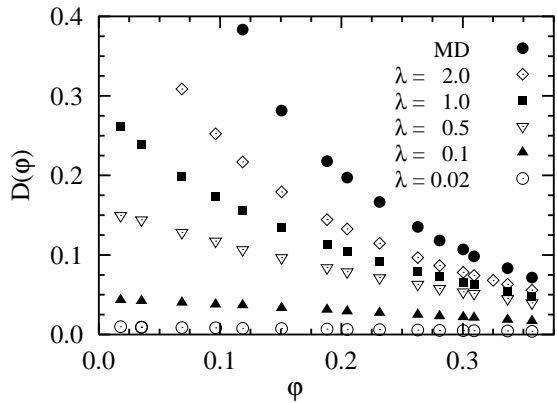,angle=-90,width= 8cm}
\caption{Diffusion coefficient as a function of the volume fraction
of colloidal particles dispersions interacting with a solvent 
represented with MPCD at different collision times. 
For comparison also plotted the MD results of Fig.~\ref{fig:coll_md}.}
\label{fig:coll_hyb}
\end{figure}

Simulation results with the hybrid method are shown in
Fig.~\ref{fig:coll_hyb}.  The simulations presented here are performed
with rotation angle $\alpha=130$, fluid number density $\rho=5$,
and mass $M=5m$ of the colloidal particle. We vary the mean free
path between $\lambda=0.02$ and $\lambda=2.0$. 

In the limit of very small volume fractions, the repulsive interactions
between colloids are negligible, and the colloidal dispersion will 
behave as the dispersion of heavy point-like particles presented in 
Secs.~\ref{sec:single_heavy} and \ref{sec:heavy_dispersion}.
In this limit, we know from Eq.~(\ref{d_M}) that the diffusion
coefficient $D(0)$ increases with 
the mean free path $\lambda$ (with $D(0)\sim \lambda$ in the molecular-chaos
approximation).
The decrease of $D(\varphi)$ with decreasing $\lambda$ displayed in
Fig.~\ref{fig:coll_hyb} arises then as a natural
consequence. Furthermore, 
the MPCD interactions of the colloid particles with the fluid imply that
the self-diffusion coefficient at small densities does not diverge 
but goes to finite value dictated by Eq.~(\ref{d_M}).

For small but finite volume fractions, in the case of large values of
$\lambda$, we observe a behavior reminiscent of the $\varphi^{-1}$
decay of hard-sphere gases, instead of the linear decrease expected
from Eq.~(\ref{dcoll.hyd}). This can be understood since, in the limit
of very large mean free paths, the colloids will essentially interact
with each other rather than with the solvent. This behavior is not 
seen in experiments of colloidal dispersions, because the diffusive 
length scale is typically much smaller than the diameter of the particles.  

\begin{figure}[ht]
\epsfig{file=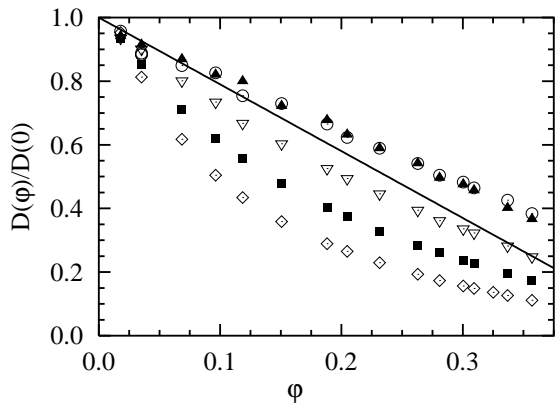,angle=-90,width= 8cm}
\caption{Dependence of the normalized diffusion coefficient on the
volume fraction $\varphi$ of colloidal particles. The same data
are shown as in  Fig.~\ref{fig:coll_hyb}. The normalization factor 
$D(0)$ is obtained by extrapolation of the data to 
zero volume fraction.
The solid line corresponds to the hydrodynamic prediction in 
Eq.~(\ref{dcoll.hyd}).}
\label{fig:coll_hyb.norm}
\end{figure}

Therefore, the appropriate parameters for the modeling of colloidal
dispersions have again to be chosen in the collective regime.
In Fig.~\ref{fig:coll_hyb.norm}, the normalized diffusion coefficient 
is shown, where $D(0)$ is extrapolated from the simulated data.
The linear behavior in Eq.~(\ref{dcoll.hyd}) is indeed observed for the
smallest values of the mean free path, $\lambda =0.02$ and
$\lambda=0.1$, within the accuracy of the simulations.
Thus, we find that in order to obtain the theoretically predicted
behavior (\ref{dcoll.hyd}) from simulation of the MD - MPCD hybrid model,
small values of the mean free path and large values of the
rotation angle $\alpha$ are required, i.e. parameters in the 
collective regime. 

However, an almost identical dependence of the diffusion coefficient
on the volume fraction is predicted theoretically in the absence of
hydrodynamics interactions.
In order to investigate this point in more detail, we have performed 
simulations of a hybrid
model similar to the one presented here, but with a completely {\em
Brownian solvent}. One way of transforming a MPCD fluid in a Brownian
solvent has been introduced by Kikuchi {\em et al.} \cite{yeo02}, where 
the velocities among all the fluid particles are randomly interchanged 
after each MPCD collision
step. We propose an alternative method which does
not consider any solvent particles. Instead, at
every $h/\Delta t$ steps the MD dynamics is interrupted for a
rotation of the (full) velocity of each colloid around a random axis by and
angle $\alpha$.  In this case,
the diffusion coefficient at zero volume fraction $D(0)$ is given by
Eq.~(\ref{d_M}) but the decorrelation factor is $\gamma\equiv\gamma_\alpha$ 
with $\gamma_\alpha$ of Eq.~(\ref{gamma.a}). The simulation results of 
the dependence of the diffusion coefficient on the volume fraction are 
quite similar to those displayed in Fig.~\ref{fig:coll_hyb.norm}. 
The data for $D(\varphi)$ follow a linear decay only for very small 
values of $\lambda$,
where the friction is large and $D(0)$ is small enough to represent a
fluid. For large values of $\lambda$, $D(\varphi)$ has a concave shape,
reminiscent of the $\varphi^{-1}$ behavior of gases,  
similarly as observed for the hydrodynamic simulations. 

Simulations with a similar hybrid method of a {\em two-dimensional} 
colloidal suspension have been reported by 
Falck {\em et al.} \cite{fal04}.  In the majority of the presented 
results, they consider excluded-volume
interaction among colloids but not between colloids and solvent
particles. They measure an apparent tracer diffusion coefficient 
(since the diffusion coefficient in two dimensions 
diverges with increasing system size) of the colloids for different 
concentrations. Three different Schmidt numbers are studied. 
A similar trend in the data is observed as in our simulations:
the normalized diffusion coefficient
increases with increasing Schmidt number, in particular for 
intermediate values of the volume fraction $\varphi$. 
However, our interpretation is different. While Falck {\em et al.}
attribute this effect to hydrodynamics, we believe that it is due 
to the crossover from gas-like to diffusive behavior of the 
colloidal dynamics.

In summary, our hybrid model describes the dynamics of a  
dispersion of hard-sphere colloids very well 
in the {\em collective regime of the solvent}.  
In the hydrodynamic interaction, only the leading
contribution for large distances is included in our model.  This
implies that lubrication forces between neighboring particles at short
distances, as well as the coupling between rotational degrees of
freedom, are neglected. We conclude from the very weak dependence of
our results for the normalized diffusion coefficients on the mean free
path, which controls the strength of the hydrodynamic interaction,
that our model works very well for not too concentrated colloidal
dispersions.

\section{Summary and Conclusions}

In this paper we have performed a detailed analysis of the
hydrodynamic properties of a fluid simulated with MPCD. We identify
two hydrodynamic regimes in terms of the parameters of 
the MPCD algorithm. The {\em particle regime} is characterized by 
dynamical properties being closer to those of a gas than to those of a
liquid. The Schmidt number is small and the dominant transport
mechanism is kinetic transport. This is the regime obtained for
large values of the collision time and/or small values of the rotation
angle. The second and more relevant regime for fluid simulations is 
the {\em collective regime}. In this regime the Schmidt 
number is large and 
collisional transport dominates over kinetic transport --- this 
characterizes liquid-like behavior. These properties are 
obtained for large values of the rotation angle and small values of
the collision time.

Different quantities have been measured in both regimes. The main
conclusion is that the diffusion coefficient shows a hydrodynamic
enhancement in the collective regime. In the study of the VACF we
observe that the behavior can be understood in both regimes as an
exponentially decay for short times and algebraic decay for long
times.  In the particle regime, a simple crossover between both
behaviors is observed while an extra intermediate behavior is
displayed in the collective regime.  This intermediate behavior of the
VACF is typically a slower than the initial exponential decay. We have
shown that the origin of this intermediate decay region is due to the
build-up of correlations by many-body collisions, which is in
conceptual agreement with the hydrodynamic behavior. The theoretical
predictions for the diffusion coefficient are based on a molecular-chaos
assumption, which gives an exponential decay of the
VACF. Consequently, a deviation from the theoretical prediction is
found in the collective regime. This deviation can be understood as a
hydrodynamic contribution to the Brownian value.

In a further step, we have investigated the differences between the
particle and the collective regime for complex fluids. We have studied
the behavior of heavy particles embedded in the MPCD fluid which can
represent solute or colloidal particles dissolved in a simple
fluid. This study demonstrates that optimal hydrodynamic coupling
occurs when the mass of the tagged particle is on the order of the
solvent mass in a collision cell.

In order to describe colloidal dispersions at finite volume fractions,
it is necessary to account for excluded volume interactions among
colloidal particles.  To this end, 
a hybrid model was studied, which combines MPCD for the solvent with 
MD simulations for the colloidal particles.  We show that  
only for parameters within the {\em collective regime} does the hybrid 
model
reproduce the proper hydrodynamic behavior. In this case, 
the results agree  well with the
theoretical calculations with and without hydrodynamic interactions,
as well as with experimental results. 

A more precise modeling of colloidal particles would require new
interactions among fluid and colloids such that fluid
particles would not freely travel through colloidal particles and
eventually angular momentum could be interchanged among them. 

In the future, it will be interesting to explore 
in which applications a more detailed description
of colloidal interactions is necessary, as compared to our simplified
model which allows more particles and larger system sizes,
and is therefore well suited to study cooperative phenomena.

\section*{Acknowledgments}
We thank G.~Vliegenthart, H.~Noguchi, D.M.~Kroll, T.~Ihle, N.~Kikuchi 
and A.~Lamura for helpful discussions.
Financial support of this work by the German
Research Foundation (DFG) within the SFB TR6, ``Physics of Colloidal 
Dispersions in External Fields'', is gratefully acknowledged.
M.R. also acknowledges partial
support from the Projects No. BFM2001-0290 and FIS2004-01934. 

\bibliographystyle{apsrev}
%\bibliography{ref_mpcd,ref_books,ref_dpd,ref_unclass,ref_ltt}

\end{document}